\documentclass[epj]{svjour}
\usepackage{latexsym}
\usepackage{graphics}
\usepackage[FIGTOPCAP, nooneline]{subfigure}

\begin{document}
\title{A statistical analysis of product prices in online markets}
\author{Takayuki Mizuno\inst{1}\thanks{\emph{Present address:} Institute of Economic Research, Hitotsubashi University, Kunitachi, Tokyo 186-8603, Japan. } \and Tsutomu Watanabe\inst{2}
%
}                     
\offprints{}          
\institute{Institute of Economic Research, Hitotsubashi University, \texttt{mizuno@ier.hit-u.ac.jp} \and Hitotsubashi University and Canon Institute for Global Studies, \texttt{tsutomu.w@srv.cc.hit-u.ac.jp}}
\date{May 15, 2009}
%
\abstract{
We empirically investigate fluctuations in product prices in online markets by using a tick-by-tick price data collected from a Japanese price comparison site, and find some similarities and differences between product and asset prices. The average price of a product across e-retailers behaves almost like a random walk, although the probability of price increase/decrease is higher conditional on the multiple events of price increase/decrease. This is quite similar to the property reported by previous studies about asset prices. However, we fail to find a long memory property in the volatility of product price changes. Also, we find that the price change distribution for product prices is close to an exponential distribution, rather than a power law distribution. These two findings are in a sharp contrast with the previous results regarding asset prices. We propose an interpretation that these differences may stem from the absence of speculative activities in product markets; namely, e-retailers seldom repeat buy and sell of a product, unlike traders in asset markets. 
} 
\maketitle
\section{Introduction}
\label{intro}
In recent years, price comparison sites have attracted the attention of internet users. In these sites, e-retailers update their selling prices every minute, or even every second. Those who visit the sites can compare prices quoted by different e-retailers, thus finding the cheapest one without paying any search costs. E-retailers seek to attract as many customers as possible by offering good prices to them, and this sometimes results in a price war among e-retailers. 

Reflecting this, prices quoted by e-retailers sometimes fluctuate wildly like asset prices. In fact, we see a lot of similarities between product prices in these internet markets and asset prices. As an example, consider a foreign exchange market, in which ``stores'' are financial institutions that quote selling and buying prices of their ``product'', foreign currencies in this case. Those who visit the forex market, namely, financial and non-financial firms who want to buy and sell foreign currencies, look for the best price among various prices quoted by various institutions. In this sense, the product and asset markets are similar at least in terms of their basic structure. More importantly, dynamic behaviors of prices in the two markets are quite similar; in particular, 
``price cascade'' sometimes occurs both in the product and asset markets. Given this understanding, we empirically investigate fluctuations in product prices by applying the methodology that is widely used in the analysis of asset prices.  

In one of the earliest studies conducted half a century ago, Mandelbrot discovered fractal properties in the prices of cotton \cite{01,02}. Following this, many researchers has started to analyze various asset prices, confirming the fractal properties in almost all of the asset prices investigated by them \cite{03,04,05,06}. The purpose of this paper is to extend this research strategy to the area of product prices in internet markets. 

Our main findings are summarized as follows. First, we find an evidence of a fractal property in the time axis for various products traded in online markets. Specifically, we find that the estimate of the Hurst exponent is close to 0.5, and that there is almost no autocorrelation in price changes, suggesting that the price process is close to a random walk. At the same time, we find that price increase/decrease is more likely to occur conditional on the multiple events of price increase/decrease, suggesting the presence of trend followers among e-retailers. Second, we find that there exists no long memory in the volatility of product price changes, which is an important difference from the previous results regarding asset prices. Third, we find that the change in a product price obeys an exponential distribution, which is in a sharp contrast with the fact that price change distributions for asset prices are typically characterized by power law.

\section{Dataset}
Our dataset is collected from ``Kakaku.com''(\textit{Kakaku} means price), one of the most popular price comparison sites in Japan, which is operated by Kakaku.com Inc. The number of e-retailers participating in this virtual market is about 1,300, and the number of products, which are identified by their barcodes, is about 300 thousand. Most of the products are consumer electronics, such as television, digital camera, personal computer, and so on. The number of users who visit the site is about 12 million per month. Our dataset contains all of the price quotes made by each of the e-retailers for each product, about 70 million price quotes in total, with second timestamp, for the period of November 1, 2006 to September 30, 2007. 

\section{Fractal property}
We start by observing a fractal property of the average price of a product across  e-retailers. Fig.1 shows price fluctuations in an LCD television, ``AQUOS LC-32GH2'' produced by Sharp, at three different time scales. The figure on the top shows price fluctuations over eleven months, from November 2006 to September 2007. A part of this figure, shown by a square, is magnified to obtain the middle one covering three months. Moreover, a part of the middle figure is magnified to obtain the bottom one covering only ten days. These figures with different time scales look quite similar, thus suggesting the presence of a fractal property in the time axis.

To investigate more on this property, we look at the  standard deviation, $\sigma(\tau)$, of the price change from $t$ to $t+\tau$, which is defined by:
\begin{eqnarray}
\sigma(\tau)\equiv \sqrt{\left\langle \left(P(t+\tau)-P(t)-\langle P(t+\tau)-P(t)\rangle \right)^2 \right\rangle}
\end{eqnarray}
where $P(t)$ represents the average price of a product, and $\langle x \rangle$ represents the average of $x$, so that $\langle P(t+\tau)-P(t)\rangle$ represents a drift term in the price process. If the price process has a fractal property, we have a scaling law as follows:	
\begin{eqnarray}
\sigma(\tau) \propto \tau^{\alpha}
\end{eqnarray}
where an exponent $\alpha$ is referred to as the ``Hurst exponent''. In particular, if the process is characterized by a random walk with drift, the exponent $\alpha$ is equal to 0.5. Fig.2 presents $\sigma (\tau)$ for AQUOS LC-32GH2: $\sigma (\tau)$ is shown on the vertical axis while $\tau$ is on the horizontal axis. We see that eq.(2) is satisfied as far as the time scale, $\tau$, is in the range of 1 minute to 3 months, indicating that the price process is close to a random walk on \textit{any} time scale between 1 minute to 3 months.

\begin{figure}
\resizebox{0.90\columnwidth}{!}{%
  \includegraphics{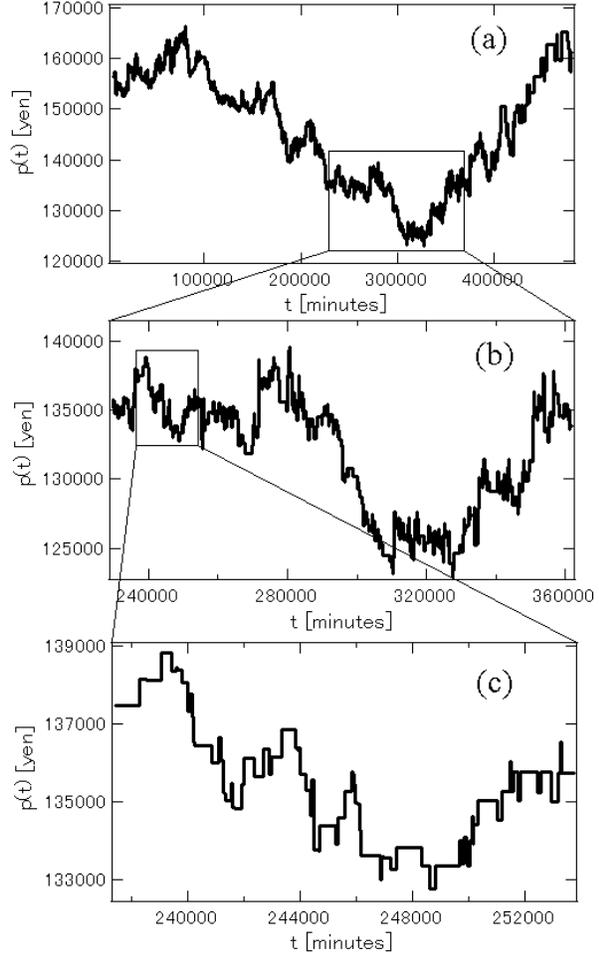}
}
\caption{The average price of a product across e-retailers at different time scales. The product is an LCD television, AQUOS LC-32GH2 produced by Sharp. The figures a, b, and c show fluctuations in the average price over eleven months, three months, and ten days, respectively.}
\label{fig:1}       
\end{figure}
\begin{figure}
\begin{center}
\resizebox{0.80\columnwidth}{!}{%
  \includegraphics{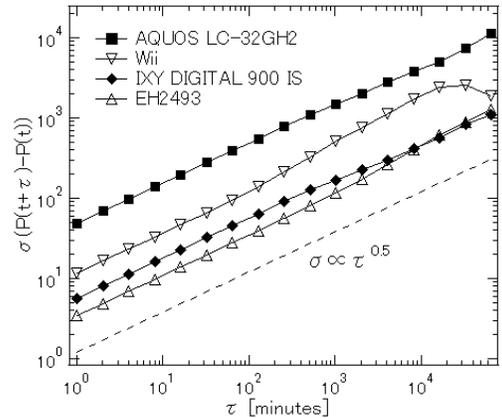}
}
\caption{The standard deviation $\sigma (\tau)$ of the price change $P(t+\tau)-P(t)$ at the time scale $\tau$ for AQUOS LC-32GH2 (shown by the square), Wii (inverted triangle), IXY DIGITAL 900IS (diamond), EH2493 (triangle). The dashed line represents $\sigma(\tau) \propto \tau^{0.5}$.}
\label{fig:2}  
\end{center}     
\end{figure}

We estimate a Hurst exponent for other products, including the digital camera, IXY DIGITEL 900IS produced by Canon, skin-care equipment, EH2493 produced by Panasonic, and a game console for Wii produced by Nintendo.  Fig.2 shows that eq.(2) is satisfied for those products as well and that the price processes for those products are close to a random walk with drift.

Previous studies about various asset prices found that eq.(2) is satisfied by those price processes, and that $\alpha$ is equal to 0.5 \cite{09,10}. This property for asset prices allows researchers and practitioners to predict price diffusion, therefore being regarded useful in conducting risk management. Similarly, our results for product prices suggest that one may use this property in order to compute a theoretical value for the appropriate buying (cost) price, thereby contributing to risk management for producers and retailers.

\section{Autocorrelation function and up-down analysis of price changes}
Fig.3 shows one-hour price changes for AQUOS LC-32GH2 over the period of November 2006 to July 2007, indicating that the price goes up and down quite wildly on this time scale.  To investigate more on this wild swing, we calculate an autocorrelation in the price change, which is defined by:
\begin{eqnarray}
\rho(T)\equiv \frac{\left\langle \Delta P(t+T) \Delta P(t)\right\rangle-\left\langle \Delta P(t) \right\rangle \left\langle \Delta P(t+T) \right\rangle}{\sigma^2}
\end{eqnarray}
where $\Delta P(t) \equiv P(t+1\mbox{hour})-P(t)$. The result, which is shown in Fig.4, indicates that there is almost no autocorrelation in the price change on this time scale, implying again that the price process is close to a random walk.

When one evaluates the relationship between $\Delta P(t)$ and $\Delta P(t+T)$ by an autocorrelation function, one compares $\Delta P$ at the two points in time ($t$ and $t+T$), and completely ignores what happens between $t$ and $t+T$. However, the events that occurs in between, like $\Delta P(t+T-1)$, could have an additional effect on $\Delta P(t+T)$ \cite{07}. To cope with such a complex correlation, a technique, often referred to as ``up-down analysis'', is adopted in the analysis of asset prices \cite{07,11,12}. 

Specifically, we now focus only on the sign of a price change by discarding information regarding the magnitude of a price change. We denote ``$+$'' when a price increases, and ``$-$'' when a price decreases. We simply ignore the event of no change in a price. Given this ``coarse graining'', we investigate a statistical property of a time series of $+$'s and $-$'s. For example, $P(-\mid--)$ represents the probability of a price decrease conditional on the occurrence of two consecutive price decreases. If the price process is a pure random walk, so that the probability of a price decrease is independent of what happened in that past, we should observe  $P(-)=P(-\mid-)=P(-\mid--) =\cdots$. 

Table.1 presents a result for an LCD television, AQUOS LC-32GH2. We see that $P(+)<P(+ \mid +)<P(+ \mid ++)<\cdots$, and that the probabilities differ from each other by more than the standard deviation. This implies that there exists a stochastic trend in the sense that a price increase is more likely to occur following the event of consecutive price increases.  A similar thing is observed for price decreases, although the difference between probabilities is not so large as compared with the case of a price increase. These results could be explained at least partially by the herding behaviors among e-retailers, or the presence of strategic complementarity in e-retailers' price setting, as has often been observed in asset markets \cite{07}.

\begin{figure}
\resizebox{0.8\columnwidth}{!}{%
  \includegraphics{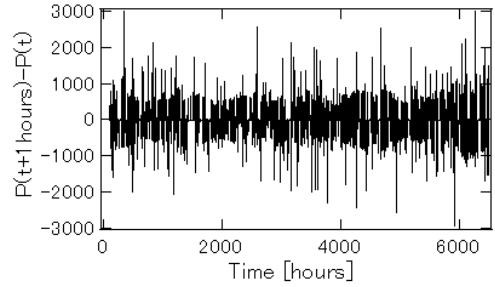}
}
\caption{One hour price changes for an LCD television.}
\label{fig:3}       
\end{figure}
\begin{figure}
\resizebox{0.8\columnwidth}{!}{%
  \includegraphics{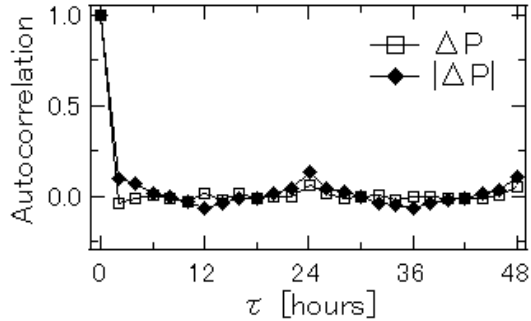}
}
\caption{Autocorrelation in the price change (shown by the square) and in the price volatility (shown by the diamond).}
\label{fig:4}      
\end{figure}

\section{Volatility and the distribution of price changes}

Traders in financial markets often pay attention to the price volatility defined by the absolute value of the price change, and researchers have estimated an autocorrelation in the volatility defined in this way for asset prices \cite{07,08}. In Fig.4, we compute such an autocorrelation in the volatility of the product price change. The figure shows that there exists no autocorrelation in the price volatility, except a few points  associated with periodical fluctuations, implying that the product price does not have a long memory in terms of the price volatility. This result is in sharp contrast with the previous results about asset prices, in which researchers have found a substantially long memory in the price volatility.

\begin{table}
\caption{Probabilities of price increase/decrease conditional on the multiple events of price increase/decrease. For example, $P(+|++)$ represents the probability of price increase conditional on the occurrence of two consecutive price increase. The standard error is defined by $1/\sqrt{n}$, where $n$ is the number of observations.}
\begin{center}
\begin{tabular}{lll}
    \hline \noalign{\smallskip}
    $P(+)$&0.24 $\pm$ 0.007\\ 
    $P(+\mid+)$ &0.35 $\pm$ 0.016\\
    $P(+\mid++)$ &0.39 $\pm$ 0.028\\
    $P(+\mid+++)$ &0.43 $\pm$ 0.046\\
    $P(+\mid++++)$ &0.44 $\pm$ 0.070\\
    $P(+\mid+++++)$ &0.50 $\pm$ 0.087\\
    $P(+\mid++++++)$ &0.57 $\pm$ 0.106\\
    $P(+\mid+++++++)$ &0.60 $\pm$ 0.115\\
   \hline
   \hline
$P(-)$ &0.76 $\pm$ 0.007\\
$P(-\mid-)$& 0.79 $\pm$ 0.008\\
$P(-\mid--)$& 0.82 $\pm$ 0.008\\
$P(-\mid---)$& 0.85 $\pm$ 0.009\\
$P(-\mid----)$& 0.86 $\pm$ 0.009\\
$P(-\mid-----)$& 0.87 $\pm$ 0.009\\
$P(-\mid------)$& 0.88 $\pm$ 0.010\\
$P(-\mid-------)$& 0.89 $\pm$ 0.010\\
\noalign{\smallskip}    \hline
  \end{tabular}
  \end{center}
\end{table}

Turning to the distributions of price changes, we show in Fig.5 the cumulative density functions (CDF) of the price change at three different time scales; $\tau=1$ minute, $\tau=100$ minutes, and $\tau=1$ week. The CDF for the positive and negative changes are shown on the panel (a) and the panel (b), respectively. Note each distribution is normalized by dividing by its standard deviation. We see that the distribution for the time scale of 1 minute has fatter tails, both at the positive and negative changes, than the standard normal distribution, which is represented by the thick dashed line. In fact, the tails of the distribution are close to those of the exponential distribution with an exponent of -1, which is indicated by the thin dashed line. However, we see less fat tails for the cases of $\tau=100$ minutes and $\tau=1$ week: in particular, the tails behave almost like the normal distribution for $\tau=1$ week. This suggests that the price change distribution converges to the normal distribution as the time scale increases.

Previous studies about asset prices have found that the tails of price change distributions for these asset prices are fatter than those of exponential distributions, and close to power law distributions \cite{06,07,08}. Put differently, a very large price change is more likely to occur in asset markets than in product markets. This is an important difference between the product and asset markets, which arises at partially due to the absence of speculative activities in the product markets. Namely, an e-retailer typically buys a product from a producer or a wholesaler and simply sells it to an end user: an e-retailer seldom repeats buy and sell activities, unlike traders in asset markets.   

\begin{figure}
\begin{center}
\subfigure[]{
\resizebox{0.8\columnwidth}{!}{%
  \includegraphics{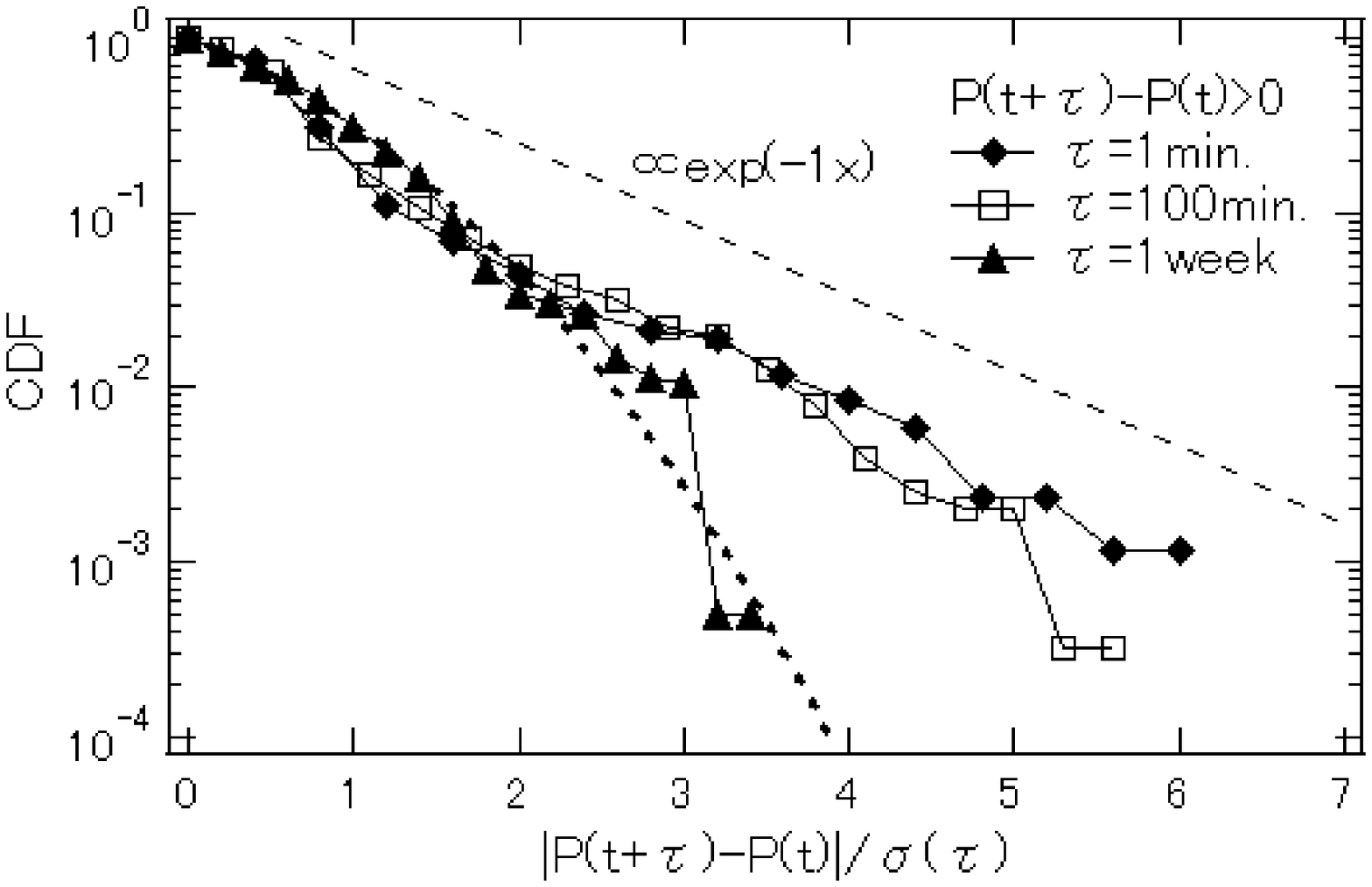}
}}
\subfigure[]{
\resizebox{0.8\columnwidth}{!}{%
  \includegraphics{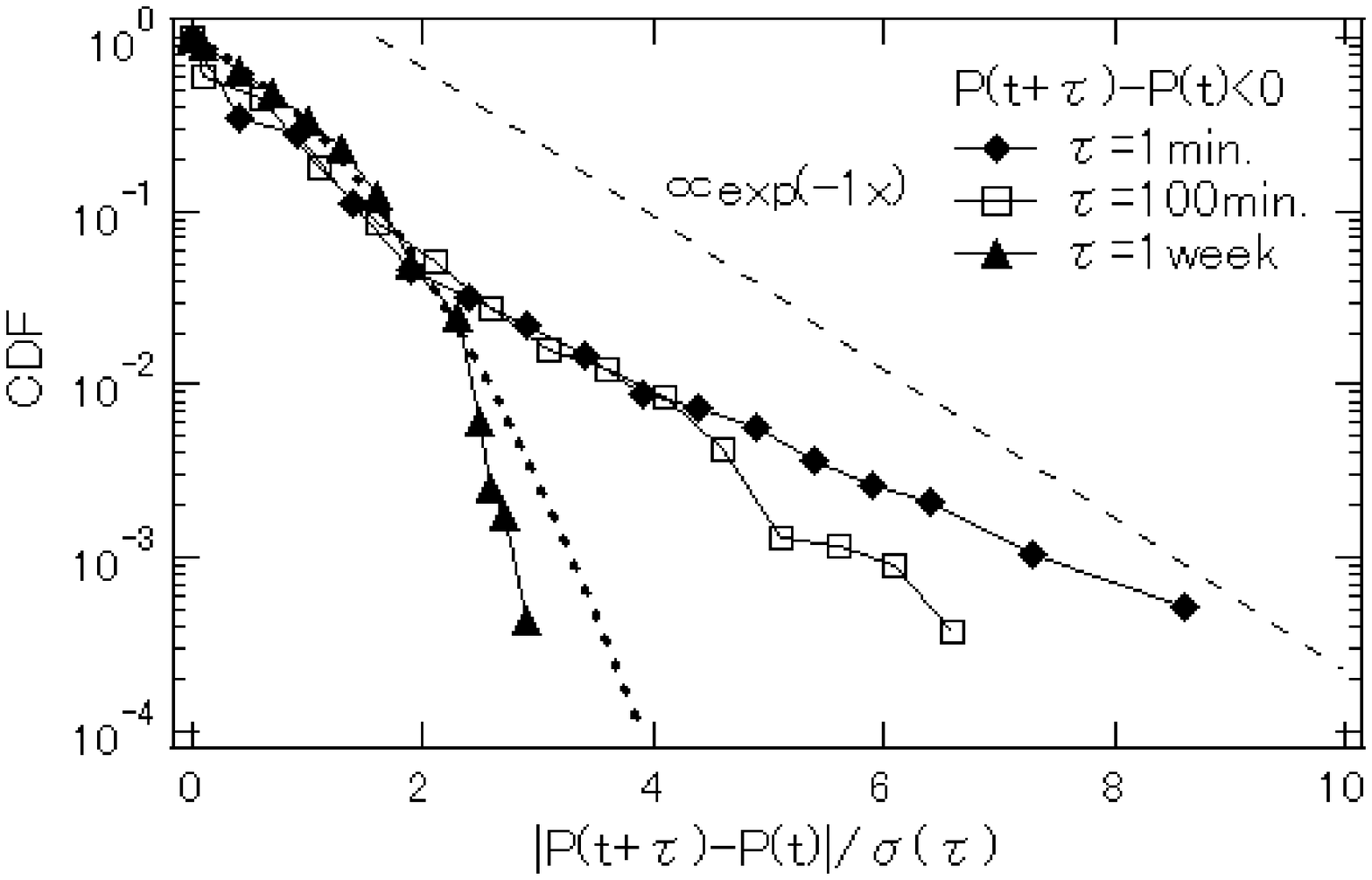}
}}
\end{center}
\caption{Semi-log plots of the cumulative density functions of the price change for $\tau=1$ minute (diamond), $\tau=100$ minutes (square), and $\tau=1$ week (triangle). The CDFs for the positive and negative price change are shown on the panel (a) and (b), respectively. Each of the distributions is normalized by its standard deviation. The thin and thick dashed lines represent an exponential distribution with an exponent of -1, and a standard normal distribution, respectively.}
\label{fig:5}      
\end{figure}

\section{Conclusion}
We have employed a unique dataset collected from a Japanese price comparison site to analyze statistical laws regarding product prices quoted by e-retailers. We have found that the Hurst exponent is close to 0.5, and there is no autocorrelation in the price change, implying that the average price of a product across e-retailers behaves almost like a random walk. However, we have also found that price increase/decrease is more likely to occur following the multiple events of price increase/decrease, suggesting the presence of trend followers among e-retailers. 

Various statistical laws regarding assets prices (stock prices and the exchange rates et al.), which were found in previous studies, has been contributing a lot to the development of risk management associated with transactions in financial markets. However, there is not so much accumulation of empirical knowledge regarding product prices, as compared with the one about asset prices. This paper is one of the first attempts to apply the methodologies that have been developed for the analysis of asset prices to the analysis of product prices. The accumulation of empirical knowledge about product prices along this line may contribute to creating new managerial technologies about production, inventory investment, pricing, and sales.

\bigskip
\bigskip
\noindent We thank Misako Takayasu and Hideki Takayasu for helpful discussions throughout the various stages of this research. We also thank Kakaku.com Inc. for providing us the dataset, and Mitsuhisa Ohdo of Kakaku.com Inc. for a detailed instruction to the dataset. This research is a part of the project entitled: Understanding Inflation Dynamics of the Japanese Economy, funded by JSPS Grant-in-Aid for Creative Scientific Research (18GS0101). T.Mizuno appreciates financial support from the Ken Millennium Corporation.

\end{document}